\documentclass[aps,prl,twocolumn,showpacs,superscriptaddress]{revtex4}

\usepackage{graphicx}% Include figure files
\usepackage{dcolumn}% Align table columns on decimal point
\usepackage{bm}% bold math
\usepackage{amsmath,amssymb}

\begin{document}

\preprint{APS/123-QED}

\title{Potential Profiling of the Nanometer-Scale Charge Depletion 
Layer in {\it n}-ZnO/{\it p}-NiO Junction 
Using Photoemission Spectroscopy}

\author{Yukiaki Ishida}
\affiliation{Department of Physics and Department of Complexity Science, 
University of Tokyo, Kashiwa, Chiba 277-8561, Japan}

\author{Hiromichi Ohta}
\affiliation{Graduate School of Engineering, 
Nagoya University, Furo-cho, Chikusa-ku, Nagoya 464-8603, Japan}
\affiliation{ERATO-SORST, JST, in Frontier Collaborative Research Center, 
S2-6F East, Mail-box S2-13, 
Tokyo Institute of Technology, 
4259 Nagatsuta-cho, Midori-ku, Yokohama 226-8503, Japan}

\author{Masahiro Hirano}
\affiliation{ERATO-SORST, JST, in Frontier Collaborative Research Center, 
S2-6F East, Mail-box S2-13, 
Tokyo Institute of Technology, 
4259 Nagatsuta-cho, Midori-ku, Yokohama 226-8503, Japan}

\author{Atsushi Fujimori}
\affiliation{Department of Physics and Department of Complexity Science, 
University of Tokyo, Kashiwa, Chiba 277-8561, Japan}

\author{Hideo Hosono}
\affiliation{ERATO-SORST, JST, in Frontier Collaborative Research Center, 
S2-6F East, Mail-box S2-13, 
Tokyo Institute of Technology, 
4259 Nagatsuta-cho, Midori-ku, Yokohama 226-8503, Japan}
\affiliation{Frontier Collaborative Research Center, S2-6F East, 
Mail-box S2-13, Tokyo Institute of Technology, 4259 Nagatsuta-cho, Midori-ku, 
Yokohama 226-8503, Japan}%

\date{\today}% It is always \today, today,

\begin{abstract}
We have performed a depth-profile analysis of 
an all-oxide {\it p}-{\it n} junction diode 
{\it n}-ZnO/{\it p}-NiO using photoemission 
spectroscopy combined with Ar-ion sputtering. Systematic core-level shifts 
were observed during the gradual removal of the 
ZnO overlayer, and were interpreted using a simple 
model based on charge conservation. Spatial profile of the 
potential around the interface was deduced, 
including the charge-depletion width of 2.3\,nm extending 
on the ZnO side and the built-in potential of 0.54\,eV. 
\end{abstract}
\maketitle
\begin{figure}
\begin{center}
\includegraphics[width=8.7cm]{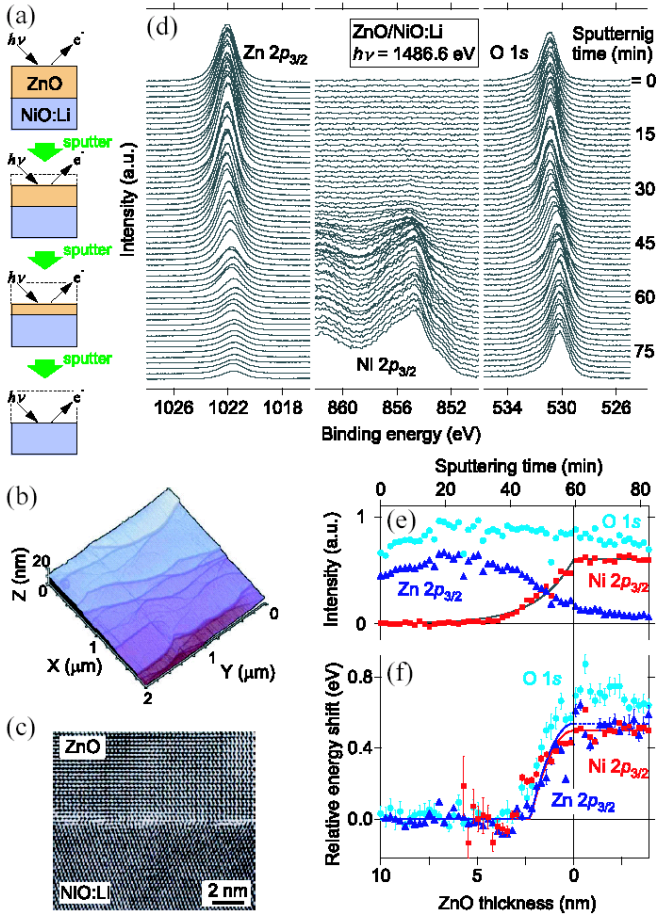}
\caption{\label{fig1} Depth-profile analysis of the 
{\it n}-ZnO/{\it p}-NiO:Li junction. (a) Schematic figure of the 
present experiment. 
(b) AFM image of a single-crystalline NiO layer 
grown on an ITO film. 
(c) HRTEM image of a 
cross-section of the ZnO/NiO heterojunction. 
(d) Variation of the Zn 2$p_{3/2}$, Ni 2$p_{3/2}$, 
and O 1$s$ core-level spectra during Ar-ion sputtering. 
(e) Variation of the core-level XPS intensities 
as a function of sputtering time. 
(f) Core-level shifts as a function of sputtering time. 
Theoretical curves for Zn 2$p$$_{3/2}$ and Ni 2$p$$_{3/2}$, 
which include 
the effect of finite photoelectron escape depth, are overlaid. }
\end{center}
\end{figure}

Oxides are considered to expand the functions of silicon-based 
devices since they show a variety of magnetic, 
electric, dielectric, and optical 
properties. Innovative oxide junction devices such as transparent field-effect 
transistors \cite{TTFT_Prins1, TTFT_Prins2, 
ZnO_TTFT_Kawai, ZnO_TTFT_Hoffman, Nomura_TFT, Nomura_FlexTFT, 
Ginley, Hosono_TCO}, 
UV-light emitters \cite{Ohta_UV, Tsukazaki}, 
and those using correlated oxides \cite{Mathews, Ahn_FET_super, 
TanakaKawai_FET, Muraoka, Katsu_Photocarrier, ZnOLSMO, Sawa, Ahn_OxideFET} 
have been reported to date. 
The characteristic width of the charge-depletion region (CDR) at an oxide 
junction interface becomes as narrow as several nanometers due to the 
generally high carrier concentrations in carrier-doped oxides 
\cite{Ahn_OxideFET}. Investigation of such a narrow CDR is nevertheless 
of primary importance since CDRs are the center of the device functions. 
In this Letter, we show a potential profile study of 
a {\it n}-ZnO/{\it p}-NiO, which is a representative and 
promising transparent 
all-oxide {\it p}-{\it n} junction diode for 
future oxide electronics \cite{Ohta_ZnONiO}. 
Instead of performing a microscopy 
around the atomically 
abrupt interface between {\it n}-ZnO and {\it p}-NiO, 
we approached the interface 
by a depth-profile analysis using x-ray photoemission spectroscopy 
(XPS) combined with Ar-ion sputtering [Fig.\ \ref{fig1}(a)]. 
Since the typical photoelectron escape depth in XPS is a few nanometers 
\cite{ProbingDepth}, depth dependent analysis with nanometer 
resolution became possible \cite{Handbook}. 

The epitaxial thin film heterostructure of 
ZnO(0001)/NiO:Li(111)/YSZ(111)/ITO(111) 
was fabricated as described elsewhere \cite{Ohta_ZnONiO}. 
Here, NiO was {\it p}-type-doped with Li to form Li$_x$Ni$_{1-x}$O. 
ZnO thickness was 10\,nm, derived from the interference fringe. 
Atomically flat surface and interface 
of the sample were confirmed by atomic force 
microscope [AFM: Fig.\ \ref{fig1}(b)] 
and high-resolution transmission electron 
microscope [HRTEM: Fig.\ \ref{fig1}(c)] observations. 
XPS measurements were performed 
using a Scienta SES-100 electron 
analyzer and an x-ray tube of Al K$\alpha$ line ($h$$\nu =$ 1486.6\,eV). 
The energy resolution was $\sim$800 meV, and the base pressure was 
better than 2$\times$10$^{-10}$\,Torr. The voltage stability was 
better than 5\,meV during the measurements, which enabled us 
to determine the energy shifts with an accuracy of $\sim$40\,meV 
albeit the rather low energy resolution of XPS. 
{\it In situ} sample etching was performed in the preparation 
chamber equipped with an ULVAC USG-3 ion gun and an Ar gas inlet. 
The energy and the incidence angle of the Ar-ion beam (defocused) 
was set 500\,eV and 85$^{\circ}$ (grazing incidence), 
respectively. During the etching, the sample was moved at 
$\sim$0.5 Hz in the vertical direction to the incidence beam 
in order to ensure homogenous etching. All the measurements 
were performed at room temperatures.

Figure \ref{fig1}(d) shows a series of core-level XPS spectra 
recorded during the removal of the initially 10\,nm-thick ZnO overlayer 
from the ZnO/NiO:Li/ITO/YSZ sample. The binding energies are referenced to 
the electron chemical 
potential $\mu$ of the sample, as in usual XPS experiments. 
In the initial stage of 
sputtering (for sputtering time $t \lesssim$ 30\,min), one could see 
signals only from the ZnO 
overlayer. The line shapes and the energy positions of the Zn 
2$p$$_{3/2}$ and O 1$s$ core-level spectra hardly changed in this stage, 
which confirms the reported chemical robustness of ZnO 
against Ar-ion sputtering \cite{Kelly}. After 
$t \sim$ 30\,min, Ni 2$p$$_{3/2}$ signals from the NiO underlayer 
became visible and grew its intensity, while that of 
Zn 2$p$$_{3/2}$ gradually disappeared. 
The O 1$s$ intensity remained nearly unchanged throughout, 
since oxygens of similar densities are present both in the ZnO 
over layer and the NiO under layer. 
At $t\simeq$ 60\,min, we observed an abrupt termination 
of the growth of the Ni 2$p$$_{3/2}$ core-level intensity. 
We interpreted this point as the complete removal of the 
ZnO layer and the exposure of NiO to the vacuum. 
Then, from the initial thickness of ZnO and $t\simeq$ 60\,min, we could 
accurately determine 
the sputtering rate of ZnO to be 0.17\,nm/min and hence the bottom axis of 
Fig.\ \ref{fig1}(e) and (f). The exponential rise of 
the Ni 2$p$$_{3/2}$ intensity for $t <$ 60\,min was best described by the 
photoelectron escape depth of 1.9 nm, as shown in the theoretical 
curve in Fig.\ \ref{fig1}(e) \footnote{The total intensity of 
the Ni 2$p$$_{3/2}$ photoelectrons 
emitted from the NiO layer underneath the $d$\,nm-thick ZnO is 
$\propto\int_0^\infty e^{-\frac{d+x}{\lambda}}\,dx 
\propto e^{-\frac{d}{\lambda}}$, 
where $\lambda$ is the mean free path of Ni 2$p$$_{3/2}$ 
photoelectrons. }. 

In Fig.\ \ref{fig1}(d), one can also see that 
the core levels were shifted toward lower binding energies during 
the etching, most notably in the series of the O 1$s$ 
core-level spectra. The relative shifts of the 
core levels are plotted in Fig.\ \ref{fig1}(f). The 
fitted simulation curves (described below) for the shifts 
of O 1$s$ and Ni 2$p_{3/2}$ 
are also shown. One can clearly see 
that all the core levels suddenly started to shift toward 
lower binding energies at $t\simeq$ 45\,min or, in terms of the 
ZnO thickness $d$, at $\sim$2.5\,nm. The shift continued 
for further sputtering until the complete removal of ZnO at 
$t\simeq$ 60\,min.

\begin{figure}
\begin{center}
\includegraphics[width=8cm]{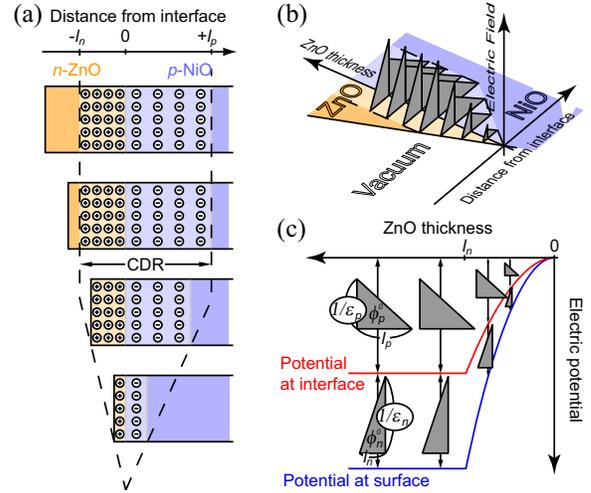}
\caption{\label{fig2} Shrink of the CDR of the {\it p}-{\it n} junction 
during the removal process of the ZnO overlayer. 
(a) Evolution of the charge distribution. 
(b) Resulting changes in the electric field distributions. 
(c) Variation in the electronic potential at the exposed 
surface and interface. 
The area of the gray triangle in (b) and (c) gives the built-in potential. }
\end{center}
\end{figure}

The sudden start of the core-level shifts with the kink at $d \sim$2.5\,nm 
would be understood if we consider the charge conservation in the CDR 
as follows. 
Figure \ref{fig2}(a) schematically shows the space-charge distributions 
within the CDR during the removal of the ZnO overlayer. 
When part of the 
CDR on the ZnO side is removed, the CDR on the NiO side should also 
shrink in order to maintain the charge neutrality. 
The associated changes in the electric field and electronic potential 
are schematically shown in Fig.\ \ref{fig2}(b) 
and (c). After the exposure 
of the CDR to the vacuum, that is, for $d < l_n$, where 
$l_n$ is the initial CDR width on the ZnO side, 
both the built-in potentials in ZnO and NiO (denoted as $\phi_n$ and 
$\phi_p$, respectively) start to diminish according to 
	$\phi_{n,p} = \phi_{n,p}^0 \times \bigl(\frac{d}{l_n}\bigl)^2$. 
Here,  $\phi_{n}^0$ and $\phi_{p}^0$ are the built-in potentials 
initially formed on the ZnO and NiO sides, respectively. 
Therefore, as shown in Fig.\ \ref{fig2}(c), the potential at 
surface and interface does not change for $d>l_n$, but 
suddenly starts to follow parabolas for 
$d<l_n$. 

Roughly speaking, the shift of Zn 2$p$$_{3/2}$ represents the variation in the 
electronic potential at the surface, since the topmost contribution from 
the ZnO layer is the largest due to the surface sensitivity of XPS. 
Similarly, the Ni 2$p$$_{3/2}$ shift represents the potential 
variation at the interface, since the contribution from the 
interfacial NiO layer is the largest. 
The simulated core-level shifts of Zn 2$p$$_{3/2}$ and Ni 2$p$$_{3/2}$ 
including the photoelectron escape depth of 1.9 nm are overlaid in 
Fig.\ \ref{fig1}(f). The best fit was obtained with the parameters 
$\phi_n^0 + \phi_p^0 =$ 0.54\,V (potential shift at the surface), 
$\phi_p^0 =$ 0.52\,V (potential shift at the interface), and $l_n =$ 2.3\,nm. 
The extra shift of the O 1$s$ peak compared to Zn 2$p$$_{3/2}$ 
and Ni 2$p$$_{3/2}$ is understood as the chemical shift 
in going from ZnO to NiO:Li. 

\begin{figure}
\begin{center}
\includegraphics[width=5.5cm]{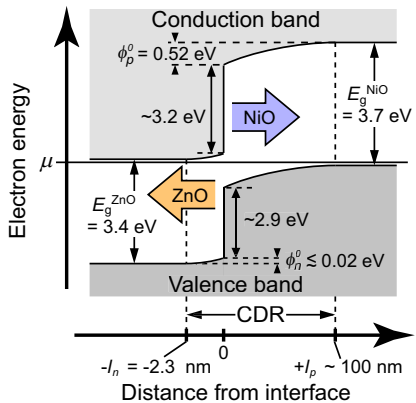}
\caption{\label{fig3}(Color online) Deduced band diagram at the 
{\it n}-ZnO/{\it p}-NiO:Li interface. Optical bandgaps of ZnO and NiO, 
$E_g^{\rm ZnO} =$ 3.4\,eV and $E_g^{\rm NiO} =$ 3.7\,eV, were adopted from 
the literature \cite{Ohta_ZnONiO}.} 
\end{center}
\end{figure}
In Fig.\ \ref{fig3}, we show the band diagram across the {\it p-n} 
junction thus deduced. 
Here, the CDR width on the NiO side $l_p =$ 1$\times$10$^2$\,nm 
was calculated using the relationship 
 $\varepsilon_n \phi_n^0 / \varepsilon_p \phi_p^0 = l_n / l_p$, 
where $\varepsilon_n = 8.59$ \cite{DielectricZnO} and 
$\varepsilon_p = 11.9$ \cite{DielectricNiO} are the 
static dielectric constant ratios of ZnO and NiO, respectively 
[see, Fig.\ \ref{fig2}(c)]. 
We have also set the conduction-band minimum 
(CBM) of ZnO and the 
valence-band maximum (VBM) of NiO:Li close to $\mu$, 
since ZnO and NiO:Li are heavily doped with electrons and holes, 
respectively. We adopted the optical gaps of ZnO and NiO of 
3.4\,eV and 3.8\,eV, respectively \cite{Ohta_ZnONiO}, as the band gaps. 
The built-in potential of 
$\phi_n^0 + \phi_p^0 =$ 0.54\,V is in good agreement with the threshold 
voltage $\sim$0.7\,V of the diode rectifying property of 
{\it n}-ZnO/{\it p}-NiO \cite{Ohta_ZnONiO}. 
The large conduction-band offset of $\sim$3\,eV is in line 
with the large energy difference in electron affinities of 
ZnO and NiO \cite{BandAlignment}. 
%This model also explains 
%the UV response above 3.4\,eV observed in transparent {\it n}-ZnO/{\it p}-NiO:Li 
%\cite{Ohta_ZnONiO}, if the UV light is absorbed 
%within the CDR by exciting the electrons across the band gaps. 

The carrier concentration of ZnO and NiO:Li can be calculated 
from the derived parameters as $N_n\sim 3\times10^{18}$\,cm$^{-3}$ and 
$N_p\sim 6\times10^{16}$\,cm$^{-3}$, respectively through the relation 
$N_n l_n = N_p l_p = \frac{2\varepsilon_0}{e}(\phi_n^0+\phi_p^0)/
(\frac{l_n}{\varepsilon_n}+\frac{l_p}{\varepsilon_p})$. 
Here, $\varepsilon_0$ and $e$ are the dielectric constant of the vacuum 
and the unit charge, respectively. 
The derived carrier concentration of ZnO is in reasonable agreement with the 
initially expected value of $N_n\sim 1\times 10^{18}$\,cm$^{-3}$ 
\cite{Ohta_ZnONiO}. 
The narrow CDR width of $l_n = 2.3$\,nm on the 
ZnO side, which corresponds to $\sim$5 unit cells of ZnO, stems from 
the high carrier concentration of ZnO. 

So far, spectroscopic studies of the abrupt 
junction regions with nanometer-to-atomic 
resolution have been performed using cross-sectional 
scanning tunneling microscopy and related techniques 
on cleaved 
junction cross-sections 
of III-V compound semiconductor heterostructures 
\cite{Yu, WolfRev, Science}. 
Thickness dependence analyses as demonstrated here will 
provide another approach in investigating the nano-scale 
electronic 
properties of the junction regions of heterostructures. 
%oxide interfaces are possible stages of new phenomena 
%\cite{Ohtomo, Okamoto} and also 
Since it is not clear how far the simple 
semiconductor physics is applicable to the interfaces of 
correlated electron systems \cite{TanakaKawai_FET, ZnOLSMO, Ahn_OxideFET}, 
%and in fact, 
%they are possible stages of new phenomena 
%\cite{Okamoto, Ohtomo}, 
firm understanding of their interfacial electronic structures 
would be necessary for further development of 
oxide junctions. 

In summary, we have performed a depth-profile analysis of 
a {\it n}-ZnO/{\it p}-NiO junction 
using core-level XPS combined with Ar-ion sputtering. 
During the gradual removal of the ZnO overlayer, 
an onset of core-level shifts was observed at a critical 
ZnO thickness $\sim$2.5\,nm. We described this behavior using a 
model based on charge conservation: 
the CDR shrinks from both sides of the junction when one side of 
the CDR is mechanically removed. We thus deduced a spatial profile 
of the potential 
around the {\it n}-ZnO/{\it p}-NiO interface, 
which is similar to the band diagram of a semiconductor {\it p}-{\it n} 
junction. The present work has demonstrated 
that the overlayer thickness dependence study can be used as a 
measure of the applicability of a semiconductor {\it p}-{\it n} junction 
picture to oxide-junction interfaces. 

The authors acknowledge Y.~Hasegawa, Y.~Yuasa, S.~Tanaka and 
A.~Tsuchiya for collaboration, T.~Mizokawa for discussion, 
K.~Nomura, K.~Okazaki, H.~Wadati and K.~Takubo for technical help. 
This work was supported by a Grant-in-Aid for Scientific Research 
in Priority Area ``Invention of Anomalous Quantum Materials" from the 
MEXT, Japan.

\end{document}